\documentclass[utf8]{FrontiersinHarvard} % for articles in journals using the Harvard Referencing Style (Author-Date), for Frontiers Reference Styles by Journal: https://zendesk.frontiersin.org/hc/en-us/articles/360017860337-Frontiers-Reference-Styles-by-Journal
\usepackage[english]{babel}
\usepackage{url,hyperref,microtype,subcaption}
\usepackage[onehalfspacing]{setspace}
\usepackage{mhchem}
\usepackage{comment}

\def\keyFont{\fontsize{8}{11}\helveticabold }
\def\firstAuthorLast{Shingledecker {et~al.}} %use et al only if is more than 1 author
\def\Authors{Christopher N. Shingledecker\,$^{1,*}$, German Molpeceres\,$^{2}$, A. Mackenzie Flowers\,$^{3}$, Deaton Warren\,$^{1}$,  Emma Stanley\,$^{1}$, and Anthony Remijan\,$^{4}$}
\def\Address{$^{1}$Department of Chemistry, Virginia Military Institute, 24450 Lexington, VA, USA \\
$^{2}$Departamento de Astrofısica Molecular, Instituto de Fısica Fundamental CSIC, Madrid, Spain \\
$^{3}$Department of Chemistry, University of Virginia, 22904 Charlottesville, VA, USA
$^{4}$National Radio Astronomy Observatory, 22903 Charlottesville, VA, USA}
\def\corrAuthor{Germ\'{a}n Molpeceres}

\def\corrEmail{german.molpeceres@iff.csic.es}

\begin{document}
\onecolumn
\firstpage{1}

\title[Aldehydes from Alcohols]{A gas-phase ``top-down'' chemical link between aldehydes and alcohols} 

\author[\firstAuthorLast ]{\Authors} %This field will be automatically populated
\address{} %This field will be automatically populated
\correspondance{} %This field will be automatically populated

\extraAuth{}% If there are more than 1 corresponding author, comment this line and uncomment the next one.
%\extraAuth{corresponding Author2 \\ Laboratory X2, Institute X2, Department X2, Organization X2, Street X2, City X2 , State XX2 (only USA, Canada and Australia), Zip Code2, X2 Country X2, email2@uni2.edu}

\maketitle

\begin{abstract}

%%% Leave the Abstract empty if your article does not require one, please see the Summary Table for full details.
\section{}

\textbf{Introduction}

\noindent
Alcohols and aldehydes represent two key classes of interstellar complex organic molecules (COMs). This work seeks to better understand their possible chemical connections, with a focus on such molecules in the sources of the star-forming region Sgr B2 (N).

\textbf{Methods}

\noindent
The gas-phase reaction between ethanol (\ce{CH3CH2OH}) and the halogens fluorine and chlorine was investigated via DFT calculations, with the goal of determining whether astrochemically viable chemical pathways leading to acetaldehyde (\ce{CH3CHO}) exist. The studied reactions were then included in an astrochemical model of Sgr B2 (N) to determine their significance under real interstellar conditions.

\textbf{Results}

\noindent
Our DFT calculations revealed that both chlorine and fluorine can react barrierlessly with ethanol to abstract a hydrogen atom. We further found that, following this initial step, the resulting ethanol radicals can undergo further reactions with atomic hydrogen, with some routes leading to acetaldehyde. Incorporation of these novel reactions in astrochemical models of hot cores suggest that they are indeed efficient under those conditions, and can lead to modest increases in the abundance of \ce{CH3CHO} during model times where gas-phase ethanol is abundant. Of the ethanol radicals included in our chemical network, we found \ce{CH3CHOH} to have the highest abundances in our simulations comparable to that of ethanol at some model times.

\textbf{Discussion}

\noindent
Overall, this work reveals a novel gas-phase `top-down'' link from alcohols to aldehydes that compliments the better studied `bottom-up'' routes involving grain-surface H-addition reactions yielding alcohols from aldehydes. Moreover, results from our astrochemical models suggest that the ethanol radical \ce{CH3CHOH} may be detectable in the interstellar medium.

\tiny
 \keyFont{ \section{Keywords:} astrochemistry, Sgr B2, DFT, astrochemical modeling, alcohols, aldehydes, halogens} %All article types: you may provide up to 8 keywords; at least 5 are mandatory.
\end{abstract}

\section{Introduction}

Hot cores, \textcolor{black}{especially those found towards the Galactic Center} in the Sagittarius (hereafter Sgr) B2 (N) complex, are known to be some of the most chemically \textcolor{black}{rich} interstellar sources \citep{belloche_complex_2013,bonfandComplexChemistryHot2019,neill_herschel_2014,loomis_detection_2013,zaleski_detection_2013}. The reason for this complexity is manyfold, though significant factors include the fact that the higher temperatures obtained during core warmup liberate existing complex organic molecules (COMs) already present in dust-grain ice mantles \textcolor{black}{out to the grain surfaces and into the gas-phase} \citep{yang_corinos_2022,mcclure_ice_2023}. This enables a comparatively brief but active period of surface chemistry due to the enhanced mobility of surface species heavier than hydrogen \citep{garrod_complex_2008,ioppolo_non-energetic_2020,qasim_experimental_2020}.

Underscoring this chemical complexity is the fact that molecules bearing most of the functional groups in organic chemistry have now been detected \citep{mcguire_2018_2018}, with many more being detected annually. \textcolor{black}{Of these, two of the most important and common are alcohols and aldehydes, having R-OH and R-CHO groups, respectively, where R could be any other part of the molecule.} One of the essential tasks in astrochemistry is investigating possible chemical connections between detected molecules. \textcolor{black}{Alcohols and aldehydes hold a crucial cornerstone in the evolution of chemical complexity in the ISM, bearing two of the most chemically ubiquitous functional groups.} To this end, a number of studies have focused on the chemical connections between alcohols and aldehydes, for example, the theoretical studies of \citet{woon_modeling_2002}, \citet{rimola_combined_2014}, \citet{das_time_2008}, \citet{song_tunneling_2017} and \citet{mondal_is_2021}, as well as the experimental work by \citet{qasim_formation_2019} and \citet{hiraoka_reactions_1998}. Most of these studies envisioned the formation of the alcohol from the aldehyde via successive reactions with atomic hydrogen on grain surfaces, which represents a ``bottom-up'' chemical link between the two. This hypothesis is supported by the easy hydrogenation of formaldehyde \ce{H2CO} to methanol \ce{CH3OH} \citep{Watanabe2002}. 

While the ``bottom-up" route is well supported for methanol and formaldehyde, the picture is more complicated for acetaldehyde and ethanol. \textcolor{black}{Recent computational works suggest that grain-surface formation routes of acetaldehyde are likely to be unfavorable \citep{perrero_quantum_2023}. This, combined with the presence of acetaldehyde in cold prestellar cores \citep{scibelli_prevalence_2020}, points to the main mechanisms of acetaldehyde formation potentially being in the gas-phase \citep{vazart_gas-phase_2020}.} Further complicating the bottom-up scenario, results from recent quantum chemical calculations by \citet{molpeceres_hydrogenation_2025} suggest that acetaldehyde, at least, is resistant to further hydrogen addition, complicating the bottom-up scenario. 

Another possible chemical link between alcohols and aldehydes is a ``top-down'' route where alcohols are converted to aldehydes through successive hydrogen abstractions. \textcolor{black}{To further investigate the role of gas-phase formation,} it is this chemical route that this work seeks to address through a combination of ab initio quantum chemical calculations and astrochemical modeling of Sgr B2(N). However, rather than assuming a series of reactions with, initially, the alcohol with H, we instead assume the initiating co-reactant are the halogen atoms Cl and F in the gas-phase. This choice was motivated in part by the observation by \citet{balucani_formation_2015} of the increased reactivity of these species compared with H. \textcolor{black}{Both HCl and HF have been detected towards Sgr B2 \citep{zmuidzinas_hcclclclc_1995,neufeld_discovery_1997}, with observed fractional abundances that exist in the ranges of $\sim10^{-10} \text{--} 10^{-9}$ relative to molecular hydrogen. The contributions of chlorine and fluorine chemistry in the context of astrochemical models have previously been studied in various interstellar environments in \citet{acharyya_gas-grain_2017}, wherein they include reactions of methanol with atomic F and Cl, which can result in either the \ce{CH3O} or \ce{CH2OH} radical. In the case of the latter, the \ce{CH2OH} can react further with a chlorine atom to produce formaldehyde in the ``top-down'' fashion, but reactions including ethanol were not investigated \citep{acharyya_gas-grain_2017}.}

The rest of this paper is structured as follows: in \S Methods we describe our approaches for both the quantum calculations and the astrochemical modeling, in \S Results \& Discussion we present the results of this work, and finally, we summarize our findings in \S Conclusions.

\section{Methods}
\subsection{Quantum Chemical Calculations}

We determined the outcome of the H-abstraction reactions with fluorine, shown in Eq. (1), and chlorine, shown in Eq. (2):

\begin{align}
    \ce{CH3CH2OH &+ F} \\
    \ce{CH3CH2OH &+ Cl}
\end{align}
using quantum chemical calculations. The first step is to determine the stationary points in each of the potential energy surface (PES) (Section \ref{sec:met:stationary}) for the different reaction channels, namely:

\begin{align}
    \ce{CH3CH2OH + F &-> CH2CH2OH + HF} \label{eq:f1} \\
    \ce{CH3CH2OH + F &-> CH3CH2O + HF} \label{eq:f2} \\
    \ce{CH3CH2OH + F &-> CH3CHOH + HF} \label{eq:f3} \\
    \ce{CH3CH2OH + Cl &-> CH2CH2OH + HCl} \label{eq:cl1} \\
    \ce{CH3CH2OH + Cl &-> CH3CH2O + HCl} \label{eq:cl2} \\
    \ce{CH3CH2OH + Cl &-> CH3CHOH + HCl} \label{eq:cl3} 
\end{align}

Later, we derive the corresponding bimolecular rate constants of each of the reaction channels (Section \ref{sec:met:kinetic}), that we later introduce in the kinetic models of \textcolor{black}{Sgr B2 (N)}.

\subsubsection{Characterization of Stationary Points} \label{sec:met:stationary}

All geometries were optimized using density functional theory (DFT) with the double-hybrid functional rev-DSD-PBEP86(D4) \citep{Kozuch2011,santra_minimally_2019,Caldeweyher2019}, in combination with the cc-pCVTZ basis set \citep{hill_correlation_2010}, which includes functions designed to account for core–valence correlation. Zero-point vibrational energy (ZPVE) corrections were computed within the harmonic approximation by numerical differentiation of the gradient. To obtain more accurate energetics, we refined the energy of each stationary point at the CCSD(T)/cc-pCVTZ level \citep{raghavachari_fifth-order_1989, bartlett_non-iterative_1990, hill_correlation_2010} using an \textcolor{black}{unrestricted Hartree-Fock (UHF) reference} wavefunction \citep{neese_approximate_2000}. All calculations, including geometry optimizations, ZPVE corrections, and single-point energy refinements, were performed correlating all electrons, i.e., without applying the frozen-core approximation. All our electronic structure calculations employ the \textsc{Orca} package (v.6.0.0) \citep{Neese2020, neese_orca_2020, neese_software_2022}.

To investigate the reactivity of each distinct hydrogen atom in ethanol, we performed \textcolor{black}{exploratory} PES scans along well-defined reaction coordinates, followed by full optimization of the resulting stationary points. All reactions were modeled (Section \ref{sec:met:kinetic}) within a general mechanistic framework involving two van der Waals complexes: one preceding the hydrogen abstraction (pre-reactive complex, PRC) and one following it (post-reactive complex). This scheme was applied to all possible hydrogen abstraction pathways, regardless of whether the corresponding transition states (TS) were submerged. The only exception was the set of reactions \ce{CH3CH2OH + F/Cl -> CH3CHOH + HF/HCl}, where PES scans were inconclusive, suggesting a barrierless abstraction mechanism with no PRC on the electronic energy surface at the rev-DSD-PBEP86(D4)/cc-pCVTZ level. To assess whether these reactions truly proceed without an activation barrier, as proposed in previous studies \citep{taketani_kinetics_2005}, we carried out nudged elastic band (NEB) calculations \citep{Henkelman2000} to confirm a strictly downhill energy profile. The NEB paths were discretized into 24 images including endpoints and computed using the \textsc{Knarr} module of the \textsc{Ash} library \citep{asgeirsson_nudged_2021, bjornsson2022ash}. 

\subsubsection{Kinetic Calculations} \label{sec:met:kinetic}

\begin{figure}[h!]
    \centering
    \includegraphics[width=0.9\linewidth]{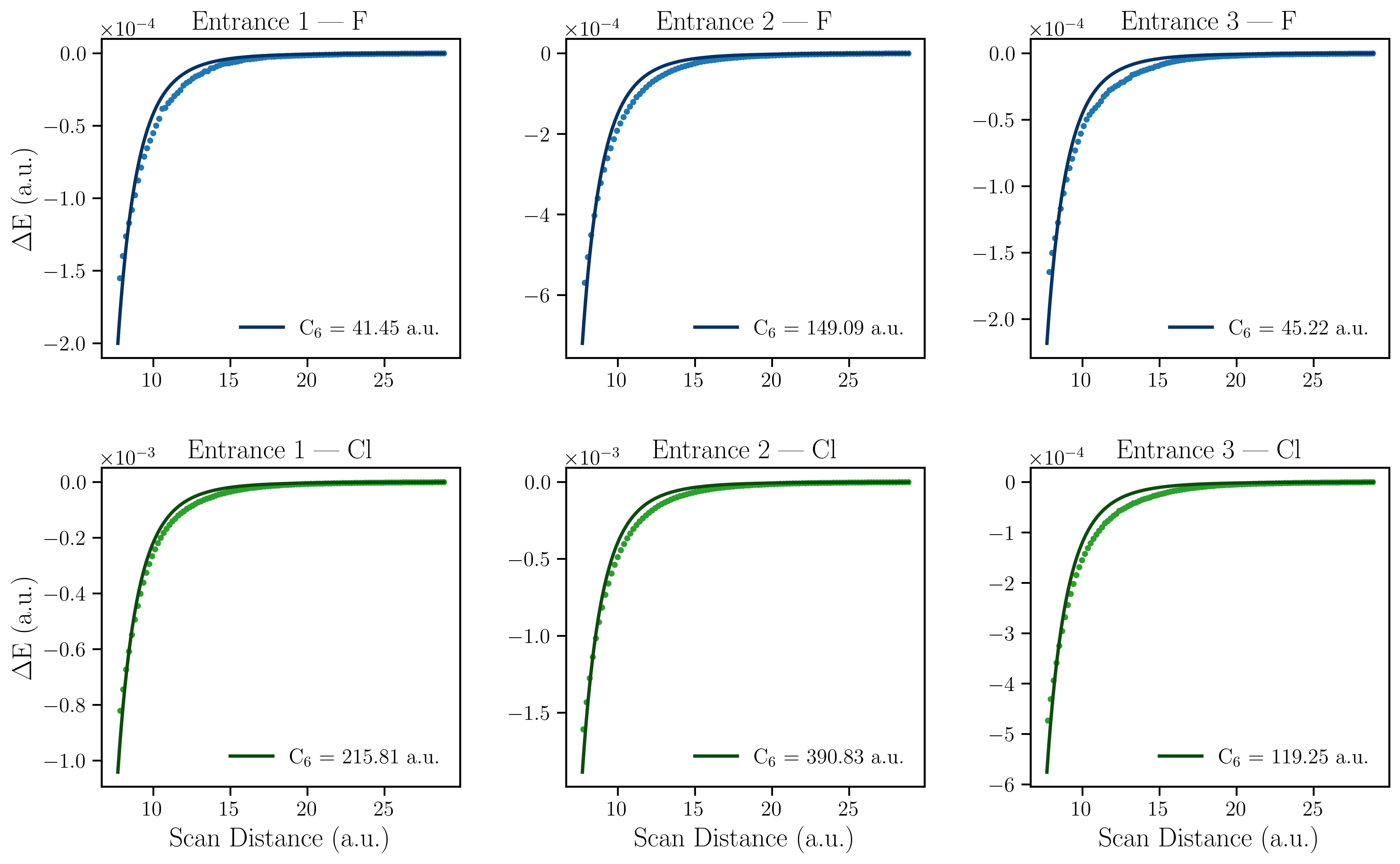}
    \caption{\textcolor{black}{Capture rigid scans leading to PRCs or VDW complexes, see text. In the legend we show the value of the capture coefficient $C_{6}$. The scans are not ZPE corrected.}}
    \label{fig:capture}
\end{figure}

We compute phenomenological rate constants for all bimolecular channels considered in this study using an ab initio transition state theory master equation (AITSTME) framework. Barrierless processes, such as capture into a pre-reactive complex (PRC) or into a product-side van der Waals (vdW) complex, as discussed above, and the escape from product vdW complexes are modeled using phase space theory \citep{Pechukas1965, Chesnavich1986}. The rigid scans for these barrierless association and dissociation channels are performed at the spin component scaled MP2 (SCS-MP2) level \citep{grimme_spincomponentscaled_2012} with the aug-cc-pVTZ  basis set \citep{kendall_electron_1992}, starting from the corresponding stationary point on the potential energy surface (\textcolor{black}{on the pre-optimized geometries using the double hybrid functional}). The scans extend from 4~\AA\ to 15~\AA\ to capture the long range asymptotic behavior of the interactions. This last set \textcolor{black}{of} calculations were carried out within the frozen core approximation. The resulting scan energies are fitted to a potential of the form $V(r) = -C / r^n$, with $n = 6$ for entrance channels and $n = 5$ for exit channels. \textcolor{black}{The choice of using a fit with $n = 5$ stems from the change in the nature of the interaction once HF or HCl are formed. While the entrance channel corresponds to an atom–molecule interaction well described by an $r^{-6}$ dependence, the exit channels are dominated by dipole–dipole interactions, better represented by an $r^{-5}$ behavior. In fact, the large dipole moment of HF (or HCl) can make some exit channels even better reproduced by an $n = 4$ dependence. Nevertheless, we retain $n = 5$ for consistency, since the kinetics of the reaction of interest are mainly governed by the capture event, for which the $r^{-6}$ fit provides a good description in all cases (Figure \ref{fig:capture}).} In cases where a barrier exists between the PRC and the product vdW complex, the interwell transition is modeled using RRKM theory, including tunneling corrections based on asymmetric Eckart transmission coefficients. To avoid numerical instabilities arising from vibrational entropy contributions, all vibrational frequencies of weakly bound complexes below 100 cm$^{-1}$ are raised to that threshold. The elementary rate constants and the resulting phenomenological rate constants for the full process are calculated using the \textsc{Mess} code \citep{Georgievskii2013}, across a temperature range of 30 to 500 K and at a residual pressure of 1$\times$10$^{-7}$ atm, low enough to prevent collisional relaxation of the PRC and product vdW complexes. Finally, in a first approximation we consider that all reactions take place in the lowest conformer of ethanol, that is, we do not consider a multistate (or multiconformer) treatment of the reaction \citep{zheng_multi-path_2012}.

The phenomenological rate constants are later fitted to a typical 3-parameter Arrhenius-Kooij formula of the type:

  \begin{equation}
    k = \alpha \left(\dfrac{T}{300\text{K}}\right)^\beta \exp{\left(-\dfrac{\gamma}{T}\right)}
    \label{eq:arrhenius}
\end{equation}

\noindent
where $\alpha$ is the pre-exponential factor for the rate coefficient, $\beta$ gives the temperature dependence, and $\gamma$ is the energy barrier. The resulting fits to this equation for reactions (3) - (8) are described below.

\subsubsection{Avenues for the improvement of the quantum chemical calculations}

\textcolor{black}{The model chemistry employed in our electronic structure calculations is sufficiently accurate for the purposes of this study. Nevertheless, several refinements could be implemented to achieve a fully quantitative description of the reactions, particularly at higher temperatures where finite-temperature effects may influence the kinetics. Possible improvements include going beyond the harmonic approximation to account for anharmonic vibrational frequencies, explicitly considering ethanol conformational effects, or introducing methodological refinements to the model chemistry, such as enlarging the basis set. The kinetic treatment could also be improved by moving beyond a one-dimensional zero-curvature tunneling correction to more accurately capture corner-cutting effects \citep{Nandi2022}. Finally, the most immediate avenue for improvement would be a more accurate description of the capture event in the barrierless channels, since capture theory is known to overestimate the capture rate constant \citep{Marchione2022, tsikritea_capture_2022}.}

\subsection{Astrochemical Modeling}

\begin{figure}[h!]
        \centering
        \includegraphics[width=0.9\textwidth]{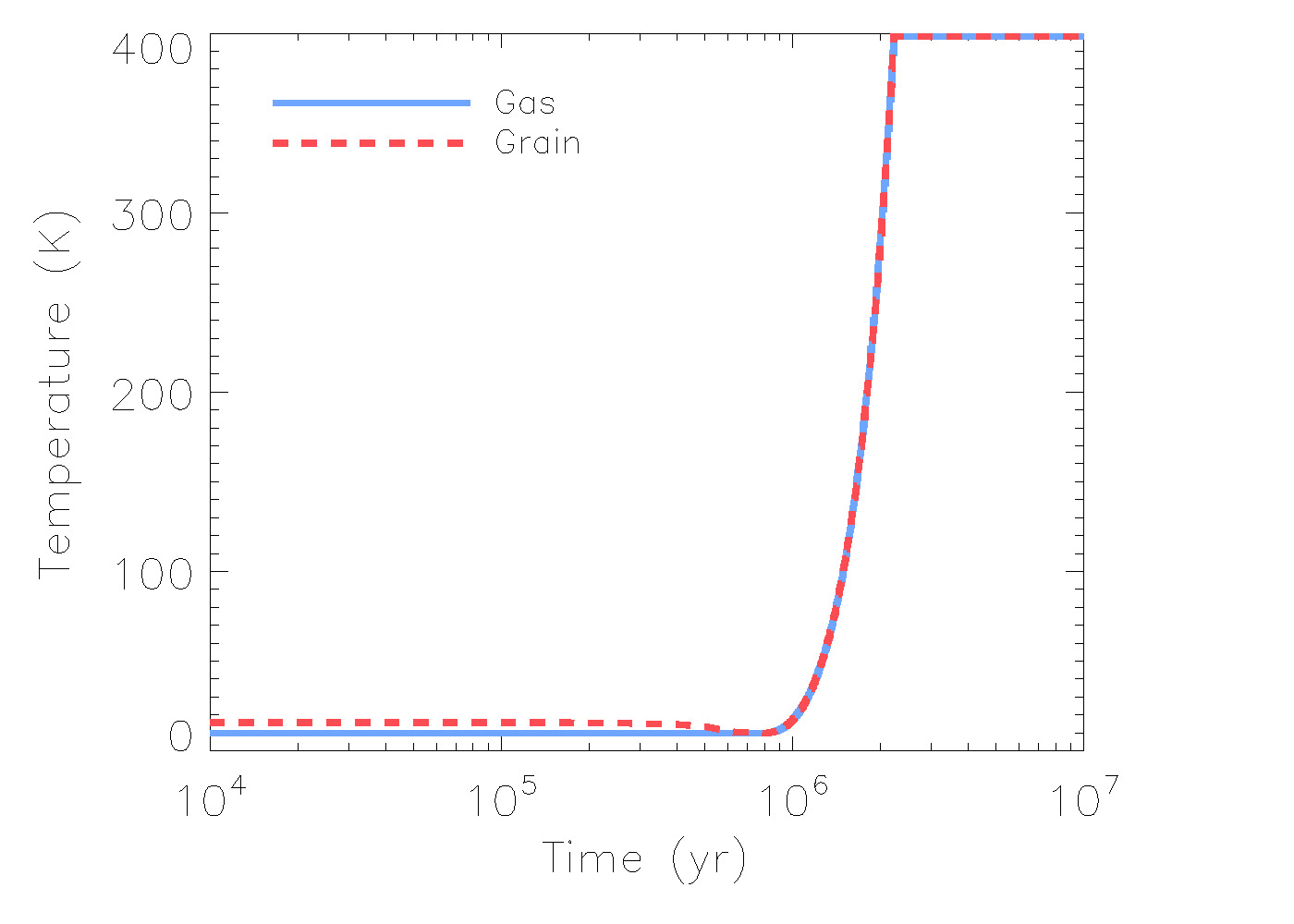}  % or .png, .jpg, etc.
        \caption{Gas and grain temperature evolution profiles.}
        \label{fig:gas_temp}
\end{figure}

\begin{figure}{h!}
        \centering
        \includegraphics[width=0.9\textwidth]{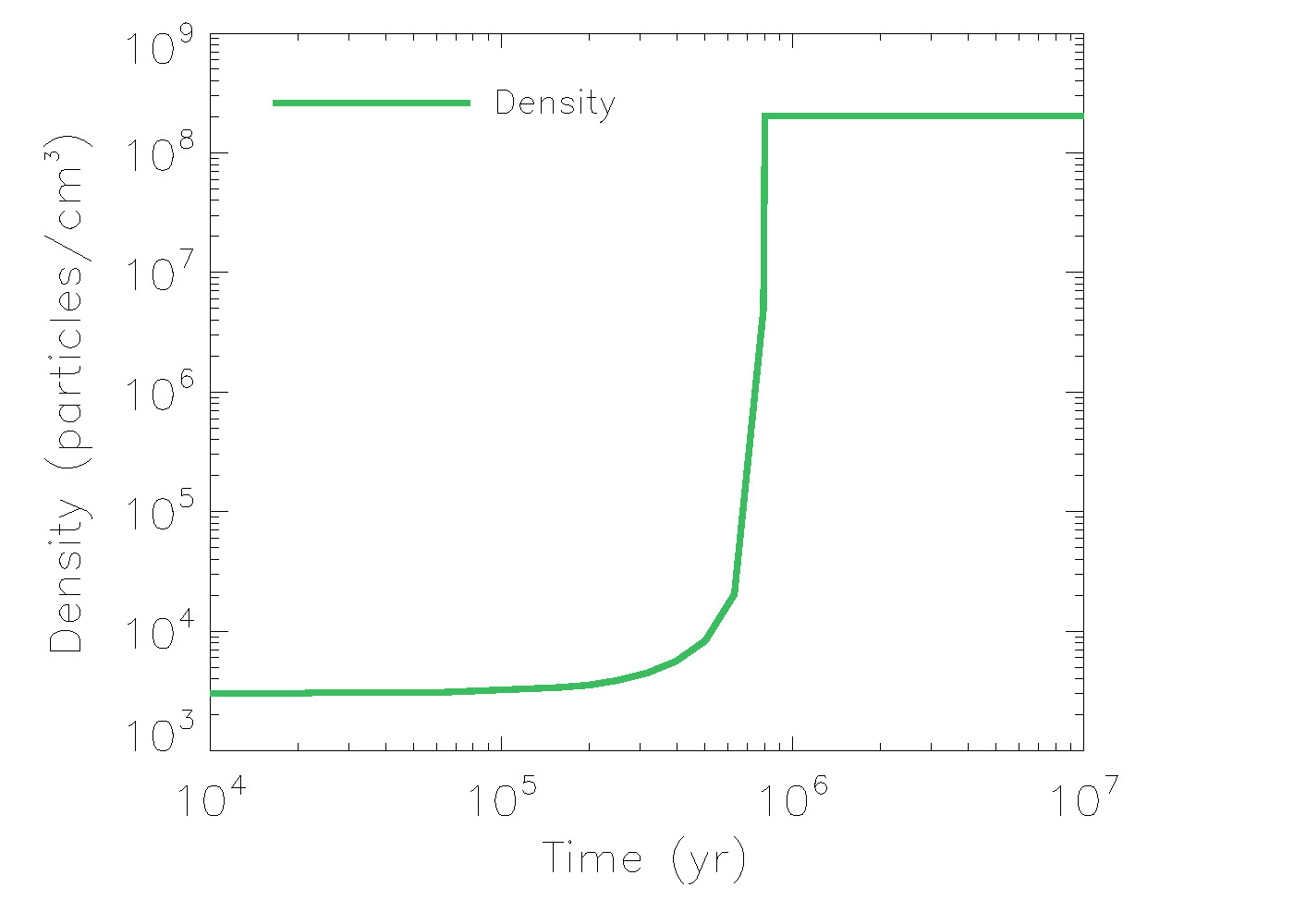}
        \caption{Density evolution profile.}
        \label{fig:density}
\end{figure}

In order to estimate the efficiency of the studied reactions under real astrophysical conditions, astrochemical models were run. For this, the \texttt{nautilus} v1.1 code was used \citep{ruaud_gas_2016}. The physical conditions used in the model, shown in Table \ref{tab:parameters}, were chosen to replicate those of Sgr B2 (N). Briefly, a three-phase (gas, grain surface, and grain bulk) three stage (collapse, warm up, and constant physical conditions) model was employed to simulate the collapse of a prestellar cloud and the formation of a hot core \citep{garrod_three-phase_2013}. The initial physical conditions at the beginning of collapse include an $A_V$ of 0.3, density of $10^3$ cm$^{-3}$. The collapse stage has a duration of $8\times10^5$ yr, at the end of which the model reaches a density of $10^8$ cm$^{-1}$, as shown in Fig. \ref{fig:density}. During the collapse, the gas temperature was held at a constant value of 10 K and the grain temperature, which starts at 15 K, becomes coupled to the gas temperature of 10 K by the end of this first stage, which can be seen in Fig. \ref{fig:gas_temp}.  A cosmic ray ionization rate of $\zeta=1.3\times10^{-16}$ s$^{-1}$ was used for all stages \textcolor{black}{to account for the enhanced cosmic ray flux in the Galactic Center, which is likely at least one order of magnitude greater than the standard value of $\sim10^{-17}$ s$^{-1}$ \citep{goto_cosmic_2013}}. 

\begin{table}[h!]
    \centering
    \caption{Physical parameters used in all astrochemical models. Note: the parameter $\alpha$ is the chemical desorption efficiency \citep{garrod_non-thermal_2007}.}
    \begin{tabular}{c|c}
      Parameter   &  Value \\ \hline
      $\zeta$ &  $1.3\times10^{-16}\;\mathrm{s}^{-1}$ \\
      UV Flux & 1 Draine \\
      $\alpha$ & $1.0\times10^{-2}$ \\
      Grain Radius & $1.0\times10^{-5}$ cm \\
      Diffusion barrier width & $1.0\times10^{-8}$ cm \\
      Surface site density & $1.5\times10^{15}$ sites/cm$^{2}$ \\
      $E_\mathrm{b}^{surface}/E_\mathrm{D}$ & 0.4 \\
    \end{tabular}
    \label{tab:parameters}
\end{table}

Following the collapse stage, a warm up occurs, during which the density and $A_V$ remain constant while, as shown in Fig. \ref{fig:gas_temp}, the gas and grain temperatures rise to $\sim400$ K over a period of $\sim10^6$ yr. Finally, after the warm up stage, the model continues with static physical conditions until a total model time of $10^7$ yr is obtained. 

Initial elemental abundances were taken from \citet{laas_modeling_2019}, and correspond to known cosmic standard elemental abundances. The full table of values is given in \citet{laas_modeling_2019}, but we note in particular an initial chlorine abundance of $2.88\times10^{-7}$ relative to hydrogen \citep{esteban_reappraisal_2004} and fluorine abundance of $3.63\times10^{-8}$ relative to hydrogen \citep{asplund_chemical_2009}. \textcolor{black}{These values differ slightly from those used in \citet{acharyya_gas-grain_2017}, which used initial abundances relative to hydrogen of $1 \times10^{-7}$ and $1.8\times10^{-8}$ of chlorine and fluorine, respectively. To determine if the differences in initial abundances will affect the overall chemistry, additional models were run with starting with initial abundances of $10^{-7}$ and $1.8\times10^{-8}$ for chlorine and fluorine, respectively.}

The base chemical network used here is that of \citet{byrne_sensitivity_2024}, developed for work by the GOTHAM (GBT Observations of TMC-1 Hunting Aromatic Molecules) project. To this was added reactions (3) - (8), in addition to a number of destruction paths for the ethanol radicals with H, OH, and \ce{NH2}. The destruction reactions with H added to our network are given in Equations (10) - (15), namely,

\begin{align}
    \ce{CH3CHOH + H -> CH3CHO + H2} \label{eq:Hd1} \\
    \ce{CH3CH2O + H -> CH3CHO + H2} \label{eq:Hd2} \\
    \ce{CH3CH2OH + H -> CH3CHO + H2} \label{eq:Hd3} \\
    \ce{CH3CHOH + H -> CH4 + H2CO} \label{eq:Hd4} \\
    \ce{CH3CH2O + H -> CH4 + H2CO} \label{eq:Hd5} \\
    \ce{CH3CH2OH + H -> CH4 + H2CO}. \label{eq:Hd6}
\end{align}

\noindent
Equations (16) - (21) show those with OH,

\begin{align}
    \ce{CH3CHOH + OH -> CH3CHO + H2O} \label{eq:OHd1} \\
    \ce{CH3CH2O + OH -> CH3CHO + H2O} \label{eq:OHd2} \\
    \ce{CH3CH2OH + OH -> CH3CHO + H2O} \label{eq:OHd3} \\
    \ce{CH3CHOH + OH -> CH3OH + H2CO} \label{eq:OHd4} \\
    \ce{CH3CH2O + OH -> CH3OH + H2CO} \label{eq:OHd5} \\
    \ce{CH3CH2OH + OH -> CH3OH + H2CO}. \label{eq:OHd6} 
\end{align}

\noindent
Finally, Equations (22) - (27) give the destruction reactions with \ce{NH2}, which are as follows:

\begin{align}
    \ce{CH3CHOH + NH2 -> CH3CHO + NH3} \label{eq:NH2d1} \\
    \ce{CH3CH2O + NH2 -> CH3CHO + NH3} \label{eq:NH2d2} \\
    \ce{CH3CH2OH + NH2 -> CH3CHO + NH3} \label{eq:NH2d3} \\
    \ce{CH3CHOH + NH2 -> CH3NH2 + H2CO} \label{eq:NH2d4} \\
    \ce{CH3CH2O + NH2 -> CH3NH2 + H2CO} \label{eq:NH2d5} \\
    \ce{CH3CH2OH + NH2 -> CH3NH2 + H2CO}. \label{eq:NH2d6} 
\end{align}

\noindent
All of the above destruction reactions were assumed to occur \textcolor{black}{barrierlessly} at the collisional rate of $10^{-10}$ cm$^{3}$ s$^{-1}$, split between the two product channels, corresponding to H abstraction (in the case of reactions leading to acetaldehyde) vs attack of the alpha carbon. \textcolor{black}{Our choice of rate coefficient is a typical approximate value for barrierless bimolecular gas-phase reactions without significant long-range forces, though follow up studies of these reactions are needed to confirm this choice, a task beyond the scope of this work. If these destruction reactions prove to be less efficient, the abundance of the intermediate radicals would be higher, which would have implications related to their detectability as described below. If the reverse is true, then the chemical connections between alcohols and aldehydes is further strengthened. Either outcome would be of astrochemical interest.}

\section{Results \& Discussion}
\subsection{Quantum Chemical Results} \label{sec:res:qc}

\subsubsection{\ce{CH3CH2OH + F}}

\begin{figure}[h!]
\begin{center}
\includegraphics[width=14cm]{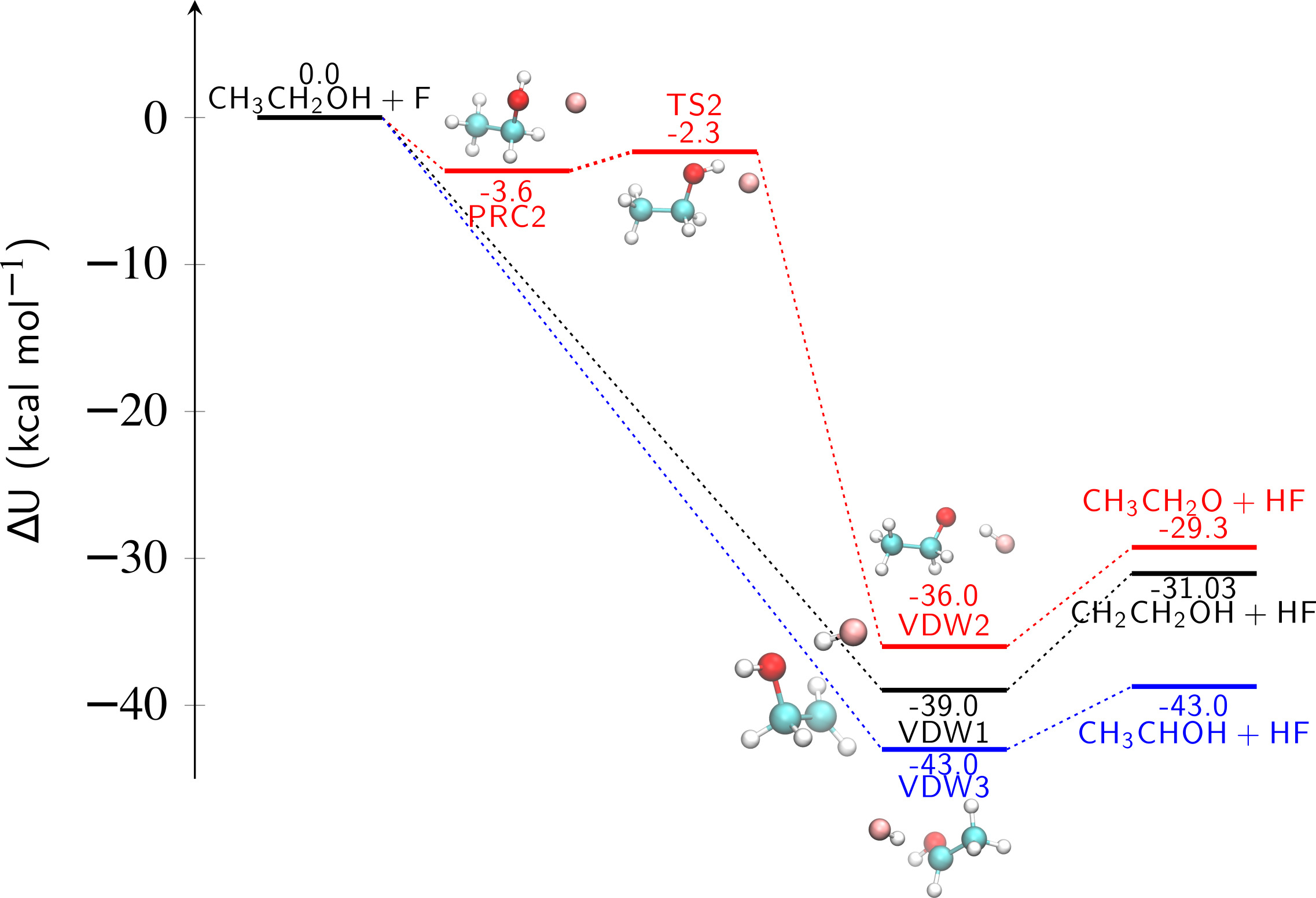} 
\end{center}
\caption{Potential energy profiles for the three abstraction channels in the \ce{CH3CH2OH + F} reaction. All energies are ZPVE corrected. }\label{fig:pes_f}
\end{figure}

The PES profiles for H abstraction by fluorine at the three inequivalent positions are shown in Figure~\ref{fig:pes_f}. A quick inspection of the graph reveals that all hydrogen abstraction reactions with fluorine (Reactions~\ref{eq:f1}, \ref{eq:f2}, and \ref{eq:f3}) are barrierless with respect to the reaction asymptote \ce{CH3CH2OH + F}. It is important to note, however, that each reaction is barrierless for a different reason. In the case of Reaction~\ref{eq:f1}, we located a TS on the electronic energy surface (that is, without ZPVE correction) at less than 1.0 kcal mol$^{-1}$. This shallow barrier disappears upon application of ZPVE. We therefore consider this reaction to be effectively barrierless, proceeding directly toward the van der Waals product complex (VDW1 in Figure~\ref{fig:pes_f}). Nevertheless, a small barrier might still be present if more accurate electronic structure methods were used to determine the stationary point energies. For Reaction~\ref{eq:f2}, the barrierless character originates from the submersion of the TS below the reactant asymptote. This is due to the stabilization of the PRC, located at approximately $-$3.6 kcal mol$^{-1}$, which consequently lowers the energy of the TS as well. Finally, Reaction~\ref{eq:f3} is the only case that is entirely barrierless, displaying a strictly downhill energy profile on the electronic PES. This behavior is confirmed by the NEB calculations described in Section~\ref{sec:met:stationary}.

\begin{figure}[h!]
\begin{center}
\includegraphics[width=10cm]{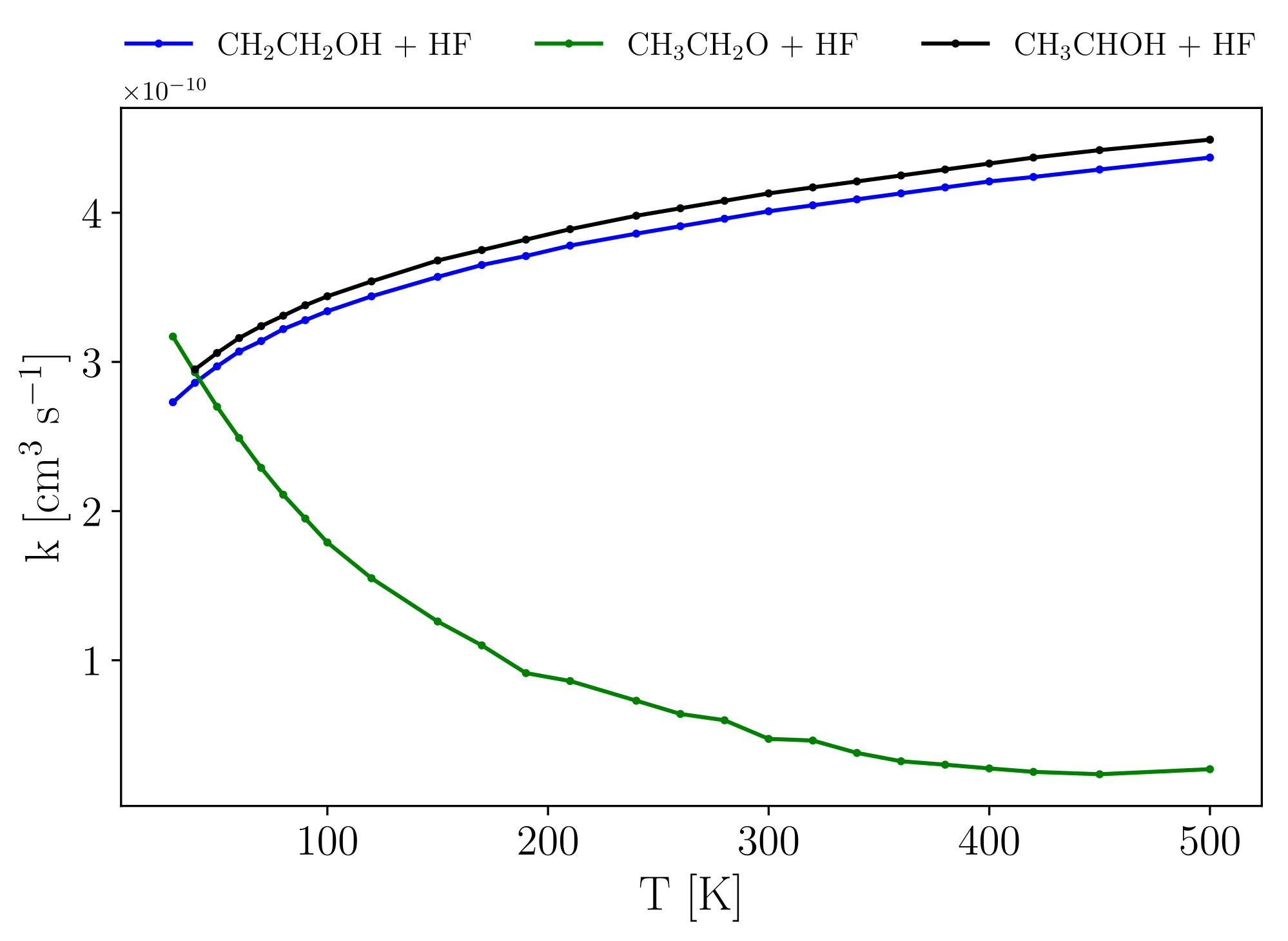} 
\includegraphics[width=10cm]{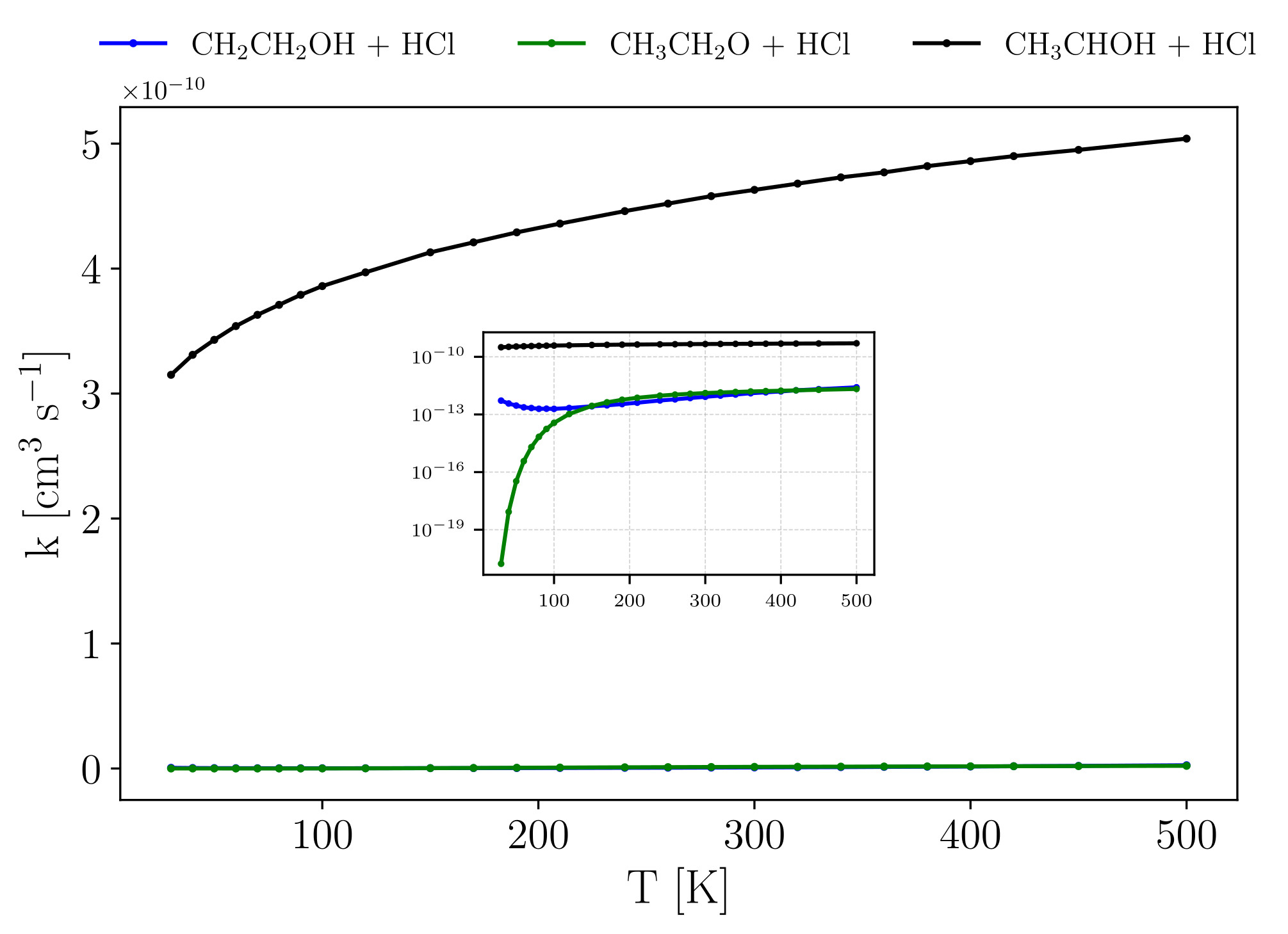}% This is a *.eps file
\end{center}
\caption{Reaction rate constants for the H-abstraction reactions of \ce{CH3CH2OH} with F (Top panel) and Chlorine (bottom panel)}\label{fig:rate_constants}
\end{figure}

The PES profiles shown in Figure~\ref{fig:pes_f} provide a useful framework to interpret the kinetics of these reactions. The corresponding rate constants are presented in Figure~\ref{fig:rate_constants} (top panel). Both Reactions~\ref{eq:f1} and \ref{eq:f3} exhibit clearly barrierless behavior, with only a minor dependence on temperature. We attribute this slight temperature dependence to the small enhancement of the final step, VDW~$\rightarrow$~Product, at higher temperatures. In contrast, Reaction~\ref{eq:f2} shows anti-Arrhenius behavior, which is typically observed in barrierless reactions where the redissociation from the PRC back to reactants becomes increasingly significant at higher temperatures.

\subsubsection{\ce{CH3CH2OH + Cl}}

\begin{figure}[h!]
\begin{center}
\includegraphics[width=14cm]{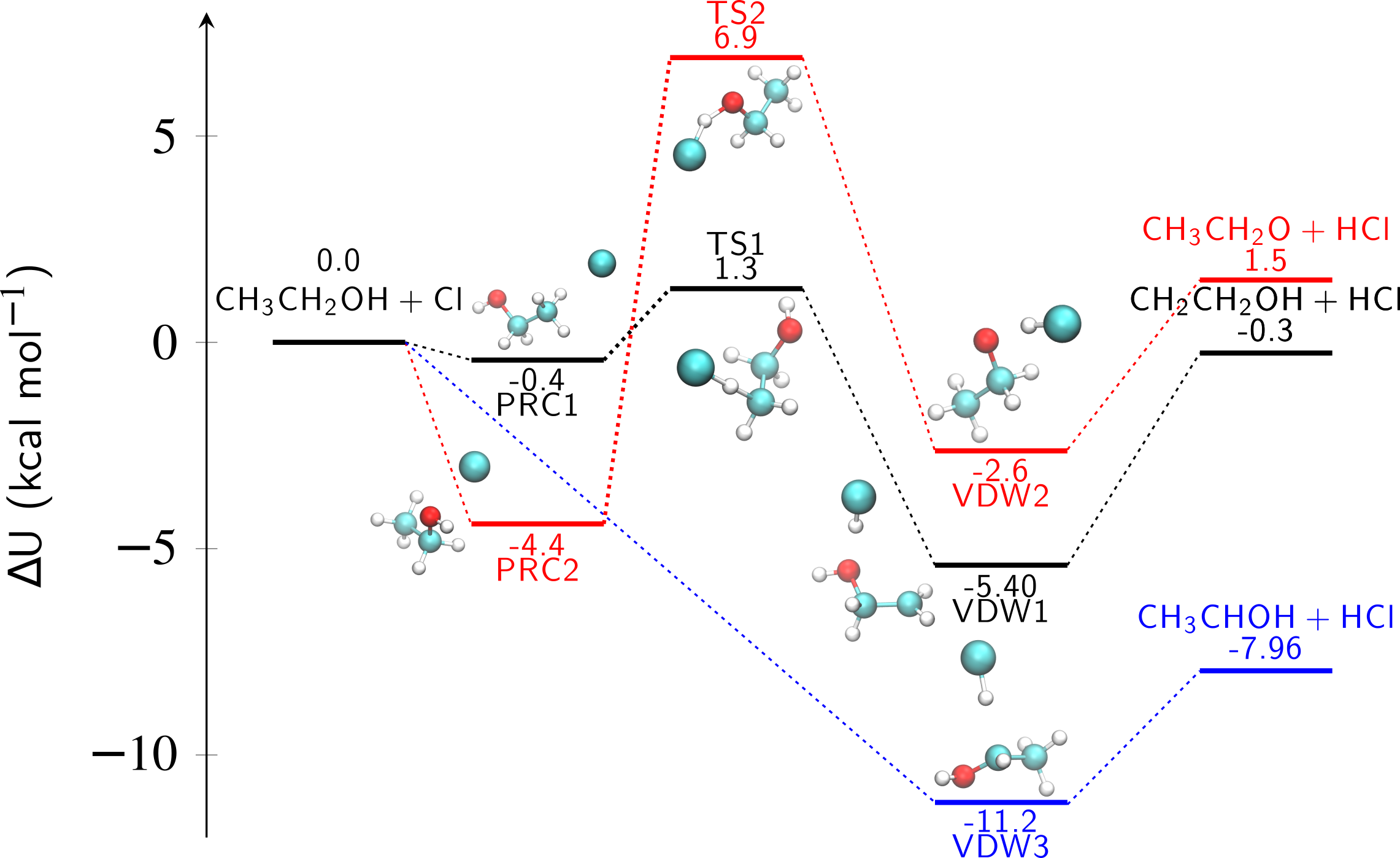}% This is a *.eps file
\end{center}
\caption{Potential energy profiles for the three abstraction channels in the \ce{CH3CH2OH + Cl} reaction. All energies are ZPVE corrected.}\label{fig:pes_cl}
\end{figure}

The \ce{CH3CH2OH + Cl} reaction is slower than its fluorine analogue. This fact follows directly from the PES cuts in Figure~\ref{fig:pes_cl}. Among the three abstraction pathways, only Reaction~\ref{eq:cl3} is truly barrierless, as confirmed by an NEB calculation, exactly as for the corresponding F reaction. The remaining pathways, Reactions~\ref{eq:cl1} and \ref{eq:cl2}, display well defined TSs above the reactant asymptote, which strongly diminish their rates. The absolute PES profile agrees closely with earlier work on this system \citep{taketani_kinetics_2005}. It reproduces both the small exothermicity of Reaction~\ref{eq:cl1} and the endothermicity of Reaction~\ref{eq:cl2}, the latter making that channel relevant only at high temperature. PRC1 (Reaction~\ref{eq:cl1}) and PRC2 (Reaction~\ref{eq:cl2}) show inverted stability with respect to their products \ce{CH2CH2OH} and \ce{CH3CH2O}. PRC2 is stabilised by H bonding between the hydroxyl group of \ce{CH3CH2OH} and the Cl atom, making it the most stable PRC in this study and even more stable than PRC2 of Reaction~\ref{eq:f2}. Such deep stabilisation raises the overall barrier for Reaction~\ref{eq:cl2}, leading to very slow kinetics. By contrast, Reaction~\ref{eq:cl1} has a much lower barrier; although slower than the barrierless channel (Reaction~\ref{eq:cl3}), its rate constants exceed those of Reaction~\ref{eq:cl2} by several orders of magnitude at low temperature. \textcolor{black}{The stabilization of PRC2, in principle, leads to competition between back-dissociation to the reactants and evolution through a roaming mechanism toward VDW3, ultimately forming \ce{CH3CHOH + HCl}. Our simplified kinetic scheme does not take roaming into account, which may result in an underestimation of the overall rate constant. However, this approximation partially compensates for the possible presence of sub-kcal mol$^{-1}$ activation barriers for reaction \ref{eq:cl3}, which would otherwise artificially enhance the calculated rate at low temperatures.}

Looking at the rate constants for the \ce{CH3CH2OH + Cl} reaction, shown in Figure~\ref{fig:rate_constants} (bottom panel), we observe that, in contrast to the case with fluorine, the reactivity is entirely dominated by the channel corresponding to Reaction~\ref{eq:cl3}. This is unsurprising given its barrierless nature. However, although irrelevant for the overall kinetics, we also find that the gap in rate constants between Reactions~\ref{eq:cl1} and \ref{eq:cl2} decreases with increasing temperature, and the two channels converge at high temperature. This unexpected trend can be rationalised by a combination of two effects. First, the greater thermal energy at high temperature makes the endothermic \ce{CH3CH2O + HCl} channel (Reaction~\ref{eq:cl2}) accessible, while it is effectively closed at low temperature. Second, the higher imaginary frequency of the TS in Reaction~\ref{eq:cl2} (2013$i$) compared to that of Reaction~\ref{eq:cl1} (1042$i$) enhances the tunneling efficiency, as this frequency can be taken as an estimate of the crossover temperature for tunneling \citep{Gillan1987}, which is significantly higher for Reaction~\ref{eq:cl2}. Nonetheless, as noted above, this behaviour does not affect the global kinetics, since both Reactions~\ref{eq:cl1} and \ref{eq:cl2} remain minor channels compared to the dominant Reaction~\ref{eq:cl3}.

\subsubsection{Comparison between reactions with fluorine and chlorine}

The comparison between the two reactions studied in this work reveals fundamental differences between H abstraction by fluorine and by chlorine, which are worth highlighting to guide further chemical interpretation. As expected from basic principles of inorganic chemistry, fluorine is overall more reactive than chlorine. However, in this case we can quantify the effect. Considering the sum of all reaction channels at low temperatures 8.85$\times$10$^{-10}$ cm$^{-3}$ s$^{-1}$, we find that the total rate constant for the fluorine system is approximately a factor 2--3 (2.8) larger than that for chlorine. The anti-Arrhenius behavior of Reaction~\ref{eq:f2}, together with a slightly faster capture efficiency in Reaction~\ref{eq:cl3} compared to Reaction~\ref{eq:f3} prevent the system from reaching the theoretical factor of three that would be expected based solely on the number of barrierless channels. It is also worth noting that while all reactions involving H abstraction by fluorine are strongly exothermic, this is not the case for chlorine, which shows much lower exothermicities and even endothermic channels at low temperatures, as in the case of Reaction \ref{eq:cl2}. This difference suggests caution when extrapolating the behavior of chlorine to other systems or when proposing new H-abstraction reactions involving chlorine.

\subsection{Astrochemical Modeling}

\begin{table}[h!]
\centering
\caption{Fitted parameters of equation \ref{eq:arrhenius} for the reactions considered in our quantum chemical calculations. \textcolor{black}{The fit is performed for the rate constants derived between 30-500 K.}}
\label{tab:arrhenius}
\begin{tabular}{lcccc}
\hline
Reaction & Label & $\alpha$ (cm$^{-3}$ s$^{-1}$) & $\beta$ & $\gamma$ (K) \\
\hline
\ce{CH3CH2OH + F -> CH2CH2OH + HF} & \ref{eq:f1} & 4.0($-$10) & 1.7($-$1) & 9.8($-$3) \\
\ce{CH3CH2OH + F -> CH3CH2O + HF} & \ref{eq:f2} & 6.1($-$11) & $-$1.4(0) & 4.5(1) \\
\ce{CH3CH2OH + F -> CH3CHOH + HF} & \ref{eq:f3} & 4.1($-$10) & 1.7($-$1) & 1.3($-$3)\\
\ce{CH3CH2OH + Cl -> CH2CH2OH + HCl} & \ref{eq:cl1} & 4.6($-$13) & 2.6(0) & $-$1.9(2) \\
\ce{CH3CH2OH + Cl -> CH3CH2O + HCl} & \ref{eq:cl2} & 7.4($-$12) & $-$4.4($-1$) & 5.2(2) \\
\ce{CH3CH2OH + Cl -> CH3CHOH + HCl} & \ref{eq:cl3} & 4.6($-$10) & 1.7($-$1) & 5.3($-$2)\\
\hline
\end{tabular}
\end{table}

\begin{figure}[h!]
    \centering
    \includegraphics[width=0.9\linewidth]{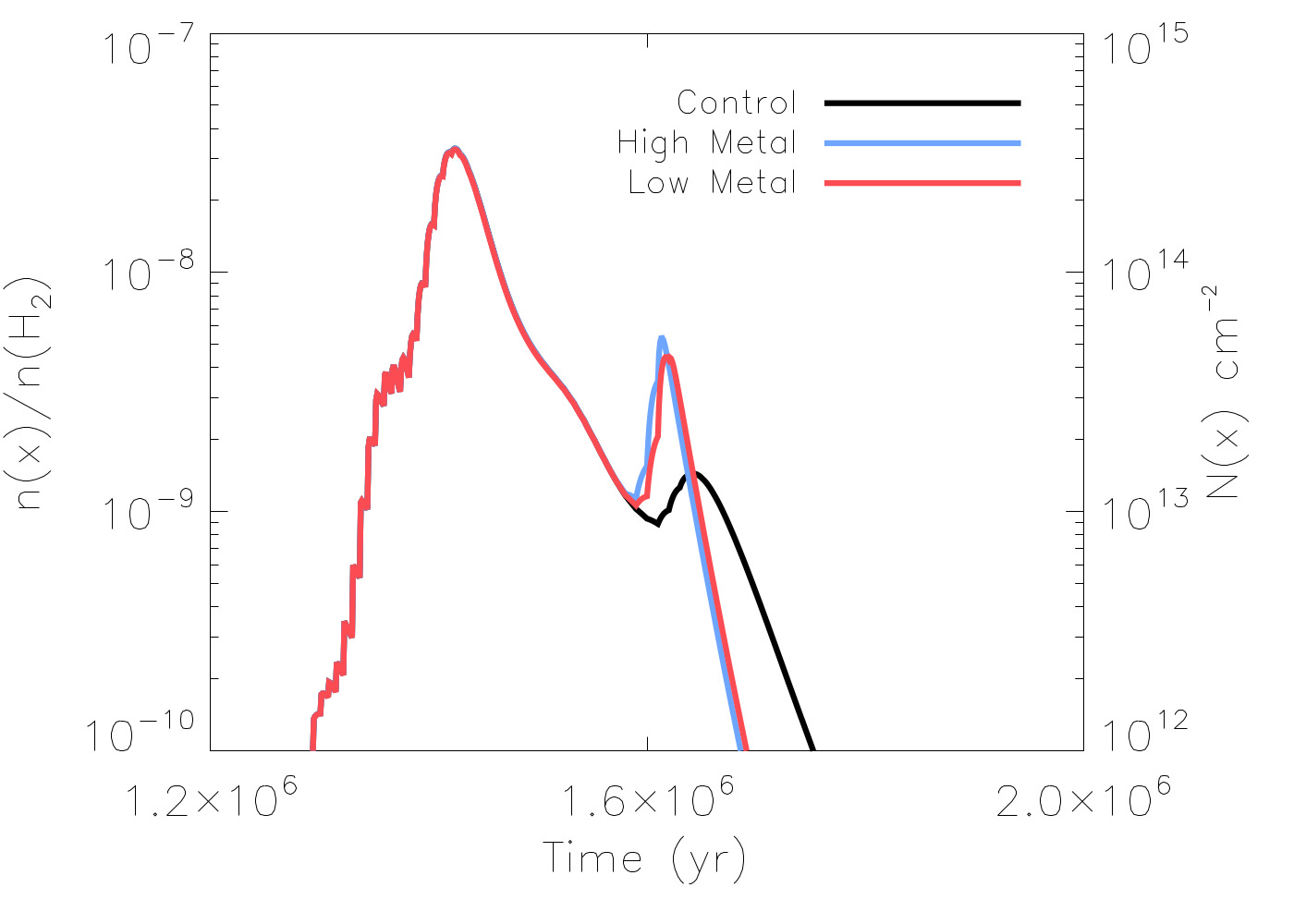}
    \caption{Abundances of \ce{CH3CHO} in our three-stage warmup model. Shown in black is the control model without the new reactions. Models including the new reactions are given in blue (high metal abundances) and red (low metal abundances). }
    \label{fig:CH3CHO}
\end{figure}

In Table \ref{tab:arrhenius} we show the values of the Arrhenius-Kooij parameters used to introduce the halogen-mediated H-abstraction reactions considered in Section \ref{sec:res:qc}. Calculated abundances for acetaldehyde (\ce{CH3CHO}) for our three-stage model of Sgr B2(N) are shown in Fig. \ref{fig:CH3CHO}. Shown in yellow are results from our ``control'' model, which uses the unmodified network of \citet{byrne_sensitivity_2024} without any of the new reactions mentioned in this work\footnote{https://zenodo.org/records/13257329}. The blue curve in the figure gives the model results using our newly expanded chemical network. A comparison of the two shows that they are largely the same, save for a noticeable increase in acetaldehyde abundance around a model time of $10^6$ yr. 

\textcolor{black}{To examine the effect of variations in initial Cl and F abundances, we ran a model using the lower initial abundances for chlorine and fluorine of, respectively, $10^{-7}$ and $1.8\times10^{-8}$ relative to hydrogen. These values have been used in previous investigations of halogen astrochemistry by, e.g., \citet{neufeld_chemistry_2005} and \citet{acharyya_gas-grain_2017}. As can be seen in Fig. \ref{fig:CH3CHO}, the reduced halogen abundances have only a minor influence on the time-dependent abundances of \ce{CH3CHO}, with the peak abundance remaining unchanged.}

\begin{comment}
Include two channels per \ce{C2H5O} radical (\ce{C2H5O + H/OH/NH2 -> CH3CHO + H2/H2O/NH3}, \ce{C2H5O + H/OH/NH2 -> H2CO + CH4/CH3OH/CH3NH2}
\end{comment}

\begin{figure}[h!]
    \centering
    \includegraphics[width=0.9\linewidth]{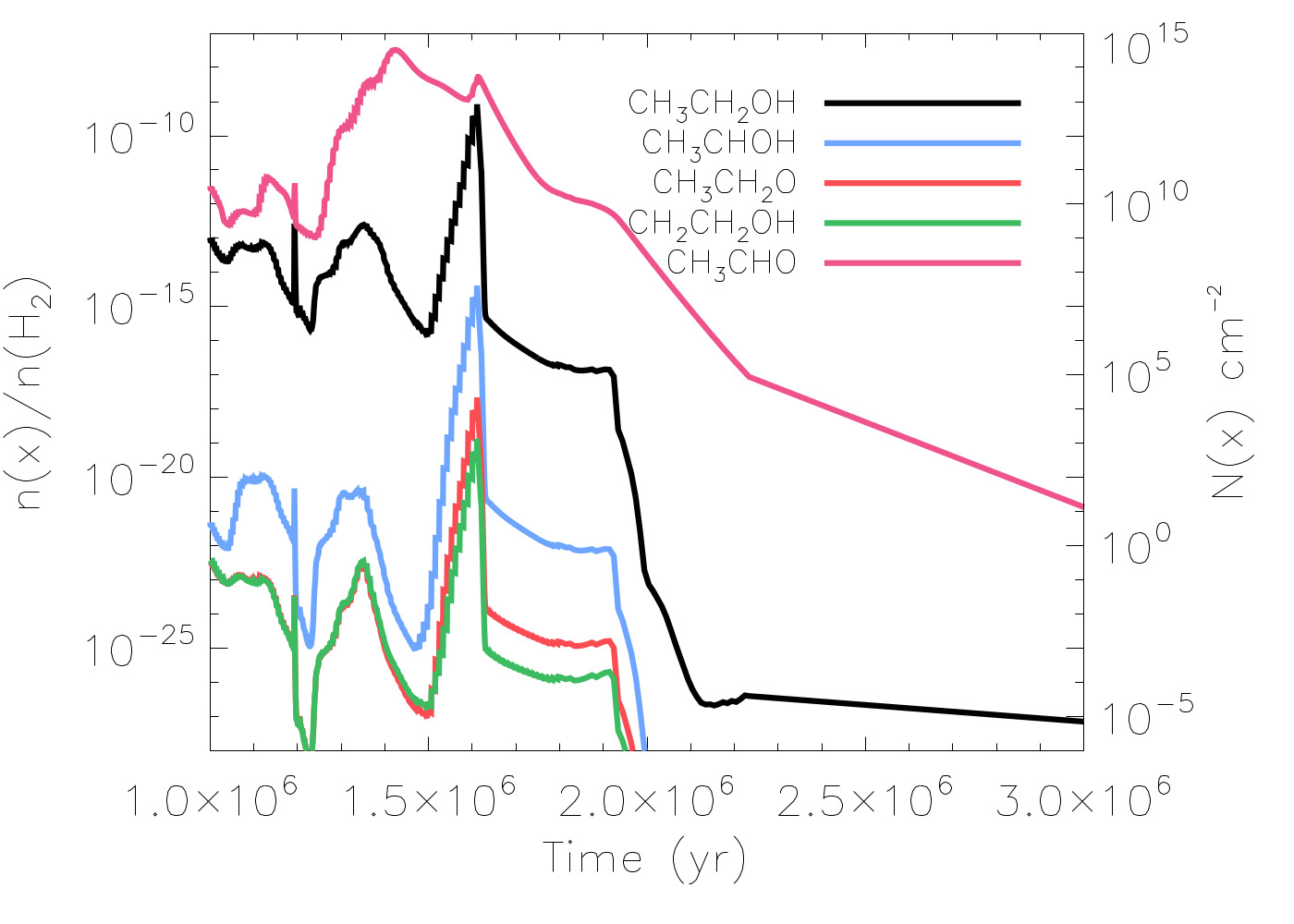}
    \caption{Abundances of \ce{CH3CH2OH}, \ce{CH3CHOH}, \ce{CH3CH2O}, \ce{CH2CH2OH}, and \ce{CH3CHO} in our new model. Results shown here are for the high metal abundance model.}
    \label{fig:multiplot}
\end{figure}

The increase in acetaldehyde abundance provides a clue as to the underlying chemistry. The likely answer is implied in Fig. \ref{fig:multiplot}. Around $10^6$ yr in the model, the gas (and dust) temperature is roughly 175 K, as shown in Fig. \ref{fig:gas_temp}. It is at this point that ethanol, which has hitherto been trapped mostly on grains, is efficiently liberated into the gas. This occurrence becomes the occasion for our new reactions to show their utility, when the desorbed ethanol begins reacting with F and Cl to produce the three ethanol radicals. Of the three, \ce{CH3CHOH} is the most abundant, reaching approximately the abundance of its ethanol precursor from a few times $10^5$ to $10^6$ yr. 

\textcolor{black}{In previous work done investigating the role of fluorine and chlorine in the chemistry of various interstellar environments (including hot cores), \citet{acharyya_gas-grain_2017} included a system of F and Cl reactions in their chemical network, including those where atomic F and Cl react with methanol. From Table \ref{tab:arrhenius} and Fig. \ref{fig:rate_constants}, we find that the temperature-dependent rate constants for reactions $3-8$ reach into the range of $\sim 10^{-13} \text{--} 5 \times 10^{-10}$ cm$^3$ s$^{-1}$. Comparing these to the chlorine and fluorine reactions with methanol, as summarized in \citet{acharyya_gas-grain_2017}, the ethanol pathways show a slight increase in reactivity. The most favourable pathway for the methanol-halogen reactions is found in the \ce{CH3OH + F} channel resulting in \ce{CH2OH}, with a rate constant of $\text{k} = 1.66 \times 10^{-10}$ cm$^3$ s$^{-1}$ \citep{jodkowski_theoretical_1999}.} 

\textcolor{black}{Despite reactions 6 and 7 not having rate constants reaching as high of values and exhibiting clear barriers, the remaining four reactions all have peak values that are at least a factor of 2 larger than their methanol counterparts \cite{acharyya_gas-grain_2017}. Furthermore, reactions 3 and 8 both share similar peak values, combined with their overall barrierless nature, this suggests that both chlorine and fluorine can efficiently contribute to the presence of the \ce{CH3CHOH} radical.}

\textcolor{black}{Our calculated} abundance presents the interesting possibility for the detection of \ce{CH3CHOH} in \textcolor{black}{interstellar environments, however, to the best of our knowledge, no laboratory spectrum for this species exists, though a recent study by \citet{williams_relative_2021} did report calculations of relative energetics. Measured laboratory spectra generally provide a more reliable basis for astronomical searches than theoretically calculated spectra. We estimate that the strongest predicted transitions may appear in the radio to centimeter range (typical of cold dark clouds) or in the millimeter to submillimeter range (typical of hot molecular cores). Given its structural similarity to both \ce{CH3CH2OH} and \ce{CH2CHOH} - two well-characterized interstellar molecules — it is plausible that the \ce{CH3CHOH} radical is also present in astronomical environments}. Completing the connection, the ethanol radicals, once formed, react with predominantly H to yield the spike in acetaldehyde abundance observed in Fig. \ref{fig:CH3CHO}. Thus, our model results show that, in times and places where gas-phase ethanol is present, our new destruction reactions with halogens represent efficient pathways linking alcohols and aldehydes.

\section{Conclusions}

In this work, we have investigated a new chemical link between alcohols and aldehydes \textcolor{black}{through use of quantum chemistry paired with astrochemical models.} Specifically, rather \textcolor{black}{than} focusing on a ``bottom-up'' formation route involving H-addition to grain-surface aldehydes, we here investigated a ``top-down'' route involving an initial reaction of an alcohol, in this case ethanol, with the halogens chlorine and  \textcolor{black}{in the gas phase}. We \textcolor{black}{find} that, in particular, the formation of \ce{CH3CHOH} by this route is efficient, \textcolor{black}{owing to barrierless reaction pathways,} and subsequent gas-phase H-abstraction could efficiently yield \textcolor{black}{acetaldehyde} in cases where gas-phase ethanol is abundant. An examination of the calculated abundances of our models including these new reactions predicts peak abundances of \ce{CH3CHOH} comparable with that of the parent species ethanol, which was initially detected in Sgr B2 by \citet{zuckerman_detection_1975}. This finding suggests that \ce{CH3CHOH} may likewise represent a \textcolor{black}{potential} target for future studies in this or similar sources, \textcolor{black}{though to our knowledge there is no spectroscopic data on this species}.

There exist many potential future directions for this work. Firstly, further study of the subsequent reactions of the ethanol radicals is warranted. Moreover, experimental and theoretical studies of this and similar systems could verify whether or not this reaction route with halogens is \textcolor{black}{accessible} for a broad range of alcohols.

\section*{Conflict of Interest Statement}
%All financial, commercial or other relationships that might be perceived by the academic community as representing a potential conflict of interest must be disclosed. If no such relationship exists, authors will be asked to confirm the following statement: 

The authors declare that the research was conducted in the absence of any commercial or financial relationships that could be construed as a potential conflict of interest.

\section*{Author Contributions}

C.N.S. contributed to the the design, organization, writing of this work, and was both the main astrochemical modeler in addition to assisting with the quantum chemical calculations. G.M. helped as well with the design, conceptualization, and writing of this work, and performed the bulk of the quantum chemical calculations. D.W. and E.S. assisted with the astrochemical modeling. A.M.F. assisted in the review and writing of the manuscript, as well as with the astronomical background. A.R. similarly contributed to the overall project conceptualization and provided relevant astronomical data and background information. 

\section*{Funding}
The National Radio Astronomy Observatory and Green Bank Observatory are facilities of the U.S. National Science Foundation operated under cooperative agreement by Associated Universities, Inc.
C.N.S., D.W., and E.S. gratefully acknowledge support through the Virginia Military Institute's Summer Undergraduate Research Institute (SURI) program. G.M acknowledges the support of the grant RYC2022-035442-I funded by MCIU/AEI/10.130 39/501100011033 and ESF+. G.M. also received support from project 20245AT016 (Proyectos Intramurales CSIC). We acknowledge the computational resources provided by the DRAGO computer cluster managed by SGAI-CSIC, and the Galician Supercomputing Center (CESGA). The supercomputer FinisTerrae III and its permanent data storage system have been funded by the Spanish Ministry of Science and Innovation, the Galician Government and the European Regional Development Fund (ERDF). 

\section*{Acknowledgments}
C.N.S. thanks A. Byrne for compiling the model inputs used as a starting point for those used in this work. 

\section*{Data Availability Statement}
The datasets generated by this work are available upon request to the corresponding author.

\begin{comment}
\section*{Supplemental Data}
 \href{http://home.frontiersin.org/about/author-guidelines#SupplementaryMaterial}{Supplementary Material} should be uploaded separately on submission, if there are Supplementary Figures, please include the caption in the same file as the figure. LaTeX Supplementary Material templates can be found in the Frontiers LaTeX folder.

\end{comment}

\bibliographystyle{Frontiers-Harvard} %  Many Frontiers journals use the Harvard referencing system (Author-date), to find the style and resources for the journal you are submitting to: https://zendesk.frontiersin.org/hc/en-us/articles/360017860337-Frontiers-Reference-Styles-by-Journal. For Humanities and Social Sciences articles please include page numbers in the in-text citations 
\bibliography{bibliography}

%%% Make sure to upload the bib file along with the tex file and PDF
%%% Please see the test.bib file for some examples of references

\end{document}